\documentclass[twocolumn,showpacs,preprintnumbers,amsmath,amssymb]{revtex4}
\usepackage{graphicx}
\usepackage{textcomp}

\begin{document}
\draft
\title{Invisibility in non-Hermitian tight-binding lattices}
  \normalsize

\author{Stefano Longhi\footnote{Author's email address:
longhi@fisi.polimi.it}}
\address{Dipartimento di Fisica, Politecnico di Milano, Piazza L. da Vinci 32, I-20133 Milano, Italy}


%
\bigskip
\begin{abstract}
\noindent Reflectionless defects in Hermitian tight-binding
lattices, synthesized by the intertwining operator technique of
supersymmetric quantum mechanics, are generally not invisible and
time-of-flight measurements could reveal the existence of the
defects. Here it is shown that, in a certain class of non-Hermitian
tight-binding lattices with complex hopping amplitudes, defects in
the lattice can appear fully invisible to an outside observer. The
synthesized non-Hermitian lattices with invisible defects possess a
real-valued energy spectrum, however they lack of parity-time
($\mathcal{PT}$) symmetry, which does not play any role in the
present work.
\end{abstract}



\maketitle

\section{Introduction}
In recent years, the subject of invisibility physics has attracted a
great and renewed interest, mainly triggered by the publication of a
few seminal papers by Pendry and Leonhardt on transformation optics
and electromagnetic cloaking \cite{Pendry,Leonhardt}, which has lead
to the first experimental observation of invisibility at microwave
frequencies \cite{sper}. Since then, a large body of works inspired
by the concepts of transformation optics has been published, and
applications to matter wave cloaking have been suggested as well
\cite{matterwaves}. An invisible object or scatter is, by
definition, an object which does not scatter any wave incident upon
it; that is, a wave which shines on the object is not reflected or
absorbed, but instead it is transmitted in such a way that it
appears to the outside observer as if there were no object present.
The concepts and methods of invisibility based on the idea of
transformation optics apply to two- or three-dimensional objects. In
one-dimensional systems, the possibility to achieve an invisible
scatter is closely related
 to the realization of reflectionless potentials. For
continuous media, this problem was investigated in a pioneering work
by Kay and Moses in 1956 \cite{Kay}, and then studied in great
detail in the context of the inverse scattering theory
\cite{inv1,inv2} and supersymmetric quantum mechanics for Hermitian
systems \cite{susy1}. The potentials obtained by such techniques,
though being transparent, are generally not invisible. This is due
to the dependence of the phase of the transmitted wave on energy,
which is generally responsible for some delay and/or for the
distortion of a wave packet transmitted across the potential
\cite{note1}.\\ The possibility of synthesizing reflectionless
potentials has been also investigated for wave scattering on a
lattice, in which wave transport occurs due to hopping among
adjacent sites of the lattice. In the mathematical literature, this
problem is solved by the inverse spectral theory of Jacobi
operators, i.e., second order symmetric difference operators
\cite{Jacobi}; in this context, Darboux transformations and the
intertwining operator technique of supersymmetric quantum mechanics
have been successfully extended to the discrete Schr\"{o}dinger
equation, with applications to the synthesis of transparent (i.e.
reflectionless) defects in Hermitian tight-binding lattices
\cite{dis1,dis2}. An optical realization of a special class of these
reflectionless potentials on a lattice has been recently proposed
for waveguide arrays and coupled-resonator structures with modulated
coupling rates \cite{Sukhorukov}, suggesting new possibilities for
pulse and beam shaping. For Hermitian lattices, such reflectionless
potentials are nevertheless not invisible because the bound states
of the lattice modify the time-of-flight of a wave packet and
generally also distorts its shape: the existence of defects in the lattice, though being transparent, could be then inferred
form simple time-of-flight measurements.\\
It is the aim of this work to show that fully invisibility of
localized defects can be realized in {\it non-Hermitian}
tight-binding lattices, which are synthesized by iterated
application of the intertwining operator technique (Darboux
transformation) to a defect-free tight-binding Hermitian lattice.
The study of non-Hermitian tight-binding lattices has received in
recent years a great attention (see, e.g.,
\cite{TB0,TB1,TB2,TB3,TB4,TB5} and references therein); such
previous studies have been mainly focused to lattices possessing
parity-time ($\mathcal{PT}$) symmetry and were framed in the context
of non-Hermitian quantum mechanics \cite{TB3,TB4,TB5,Bender},
however the possibility to realize invisibility in a non-Hermitian
lattice was not investigated in such previous works \cite{note2}.
It should be noted that the class of non-Hermtian lattices synthesized in the
present work by application of the Darboux transformation and showing the property
of invisibility are not $\mathcal{PT}$-symmetric. Nevertheless, their energy spectrum is
real-valued because they are isospectral to an Hermitian lattice. Therefore,
$\mathcal{PT}$ symmetry does not play any role in the realization of invisible defects discussed in this work.\\
The paper is organized as follows. In Secs.II and III, the
intertwining operator technique and its application to the synthesis
of tight-binding lattices with reflectionless defects are briefly
reviewed. The scattering and invisibility properties of the
synthesized lattices are discussed in Section IV; in particular, it
is shown that, as for any Hermitian lattice invisibility can never
be achieved and time-of-flight measurements can be used to reveal
the existence of defects in the lattice, in non-Hermitian lattices
with certain complex hopping rates invisibility can occur. The main
conclusions are outlined in Sec.V, whereas some mathematical details
and a possible realization of non-Hermitian lattice models based on
light propagation in optical waveguide
arrays are presented in three Appendixes.\\

\section{The intertwining operator technique for spectral engineering of tight-binding lattices}
The synthesis of reflectionless and invisible defects in a
tight-binding lattice discussed in the next sections is based on the
discrete analogs of the intertwining operator technique of
supersymmetric quantum mechanics \cite{susy1}. Extensions of the
intertwining operator technique to the discrete Schr\"{o}dinger
equation, together with the related issue of inverse scattering for
Jacobi operators, have been discussed mainly in the mathematical
literature (see, for instance, \cite{Jacobi,dis1,dis2}), however
they are not so common in the physical contexts. In this section we
thus provide a brief review of the intertwining operator technique
and its application to the problem of spectral engineering of
tight-binding lattices.\\
Let us consider a one-dimensional tight-binding lattice described by
the Hamiltonian
\begin{equation}
\mathcal{H}= \sum_{n}  \kappa_{n} \left( |n-1\rangle \langle n|+|n
\rangle \langle n-1|\right)  + \sum_n V_n |n \rangle \langle n|
\end{equation}
where $|n\rangle$ is a Wannier state localized at site $n$ of the
lattice, $\kappa_n$ is the hopping rate between sites $|n-1 \rangle$
and $|n \rangle$, and $V_n$ is the energy of Wannier state $|n
\rangle$. Note that $\mathcal{H}$ turns out to be Hermitian provided
that the hopping amplitudes $\kappa_n$ and site energies $V_n$ are
 real-valued parameters. Let us indicate by $\mathcal{H}_1$ the tight-binding Hamiltonian
defined by Eq.(1) with hopping amplitudes and site energies given by
$\kappa^{(1)}_n$ and $V^{(1)}_n $, respectively, and let us assume
that $\kappa_n^{(1)} \rightarrow \kappa>0$ and $V_n^{(1)}
\rightarrow 0$ as $n \rightarrow \pm \infty$, i.e. that the lattice
is asymptotically homogeneous and free of defects. Let
$\sigma^{(1)}=\sigma_c \cup \sigma_p$ be the spectrum of
$\mathcal{H}_1$, which comprises the continuous spectrum $\sigma_c$
(the tight-binding band $ -\kappa <E< \kappa$) and the point
spectrum $\sigma_p$. Our goal is to synthesize a new tight-binding
lattice Hamiltonian $\mathcal{H}_2$ of the form of Eq.(1), whose
spectrum $\sigma^{(2)}$ is the same as that of $\mathcal{H}_1$,
except for the addition of a new real-valued energy level $\mu_1$ in
the point spectrum, with $|\mu_1|> 2 \kappa$. To this aim, let us
indicate by $| \phi^{(1)} \rangle= \sum_n \phi_n^{(1)} | n \rangle$
a solution to the second-order difference equation
\begin{equation}
\kappa_n^{(1)} \phi_{n-1}^{(1)}+\kappa_{n+1}^{(1)}
\phi_{n+1}^{(1)}+V^{(1)}_n \phi_{n}^{(1)}=\mu_1 \phi_{n}^{(1)}
\end{equation}
with the asymptotic behavior $|\phi_{n}^{(1)}| \rightarrow \infty$
for $n \rightarrow \pm \infty$. Note that such a solution does exist
because $\mu_1$ does not belong to the point spectrum nor to the
continuous spectrum of $\mathcal{H}_1$. More precisely,
$\phi_n^{(1)}$ is given by an arbitrary superposition of two
linearly-independent solutions to Eq.(2), which behave
asymptotically as $\phi_n^{(1)} \sim \exp( \pm \omega_1 n)$ at $n
\rightarrow \pm \infty$ for $\mu_1 > 2 \kappa$, or as $\phi_n^{(1)}
\sim (-1)^n \exp( \pm \omega_1 n)$ at $n \rightarrow \pm \infty$ for
$\mu_1 < - 2 \kappa$, where $\omega_1>0$ is the root of the equation
$2 \kappa \cosh(\omega_1)=|\mu_1|$. It can be then shown by direct
calculations that the following factorization for $\mathcal{H}_1$
holds
\begin{equation}
\mathcal{H}_1=\mathcal{Q}_1 \mathcal{R}_1+\mu_1
\end{equation}
where
\begin{eqnarray}
\mathcal{Q}_1 & = & \sum_n \left( q_{n}^{(1)} |n \rangle \langle
n|+\bar{q}_{n-1}^{(1)} |n-1 \rangle \langle n| \right) \\
\mathcal{R}_1 & = & \sum_n \left( r_{n}^{(1)} |n \rangle \langle
n|+\bar{r}_{n+1}^{(1)} |n+1 \rangle \langle n| \right)
\end{eqnarray}
and
\begin{eqnarray}
r_n^{(1)} & = & - \sqrt{\frac{\kappa_n^{(1)}
\phi_{n-1}^{(1)}}{\phi_n^{(1)}}} \\
\bar{r}_n^{(1)} & = & -\frac{\kappa_{n}^{(1)}}{r_{n}^{(1)}} \\
q_n^{(1)} & = & -r_n^{(1)} \\
\bar{q}_n^{(1)} & = & - \bar{r}_{n+1}^{(1)}.
\end{eqnarray}
Let us then introduce the new Hamiltonian $\mathcal{H}_2$ obtained
from $\mathcal{H}_1$ by interchanging the operators $\mathcal{R}_1$
and $\mathcal{Q}_1$, i.e. let us set
\begin{equation}
\mathcal{H}_2=\mathcal{R}_1 \mathcal{Q}_1+\mu_1.
\end{equation}
$\mathcal{H}_2$ will be referred to as the partner Hamiltonian of
$\mathcal{H}_1$. Using  Eqs.(4-9), from E.(10) it can be readily
shown that $\mathcal{H}_2$ describes the Hamiltonian of a
tight-binding lattice [i.e., it is of the form (1)] with hopping
amplitudes and site energies $\{ \kappa^{(2)}_{n},V^{(2)}_n\}$ given
by
\begin{eqnarray}
\kappa^{(2)}_n & = & \kappa_n^{(1)} \frac{r^{(1)}_{n-1}}{r^{(1)}_n} \\
V_n^{(2)} & = & V_n^{(1)}+\kappa_{n+1}^{(1)}
\frac{\phi_{n+1}^{(1)}}{\phi_n^{(1)}}-\kappa_n^{(1)}
\frac{\phi^{(1)}_n}{\phi^{(1)}_{n-1}}.
\end{eqnarray}
Note that, owing to the asymptotic behavior of $\kappa^{(1)}_n$,
$V^{(1)}_n$ and $\phi_n^{(1)}$ at $ n \rightarrow \pm \infty$, one
has $\kappa^{(2)}_n \rightarrow \kappa$ and $V^{(2)}_n \rightarrow
0$ for  $ n \rightarrow \pm \infty$, i.e. the partner lattice
described by the Hamiltonian $\mathcal{H}_2$ is still a homogeneous
lattice without defects at $ n \rightarrow \pm \infty$. An
interesting property of the Hamiltonian $\mathcal{H}_2$ is that its
spectrum $\sigma^{(2)}$ is given by $\sigma^{(2)}=\sigma^{(1)} \cup
\{ \mu_1 \}$, i.e. it is the same as that of $\mathcal{H}_1$ plus
the additional energy level $\mu_1$ in the point spectrum. In fact,
let us indicate by $| \psi_E \rangle=\sum_n \psi_n(E) | n \rangle$ a
proper (or improper) eigenfunction of $\mathcal{H}_1$ with energy
$E$. Note that, if $E$ belongs to the point spectrum of
$\mathcal{H}_1$, $|\psi_n(E)| \rightarrow 0$ as $n \rightarrow \pm
\infty$, whereas if  $E$ belongs to the continuous spectrum of
$\mathcal{H}_1$, $|\psi_n(E)|$ remains bounded as $n \rightarrow \pm
\infty$. Since $\mu_1$ does not belong to the point spectrum of
$\mathcal{H}_1$, one has  $E \neq \mu_1$. Using the factorization
(3) for $\mathcal{H}_1$, the eigenvalue equation $\mathcal{H}_1 |
\psi_E \rangle = E | \psi_E \rangle$ reads explicitly
\begin{equation}
\mathcal{Q}_1 \mathcal{R}_1 | \psi_E \rangle=(E-\mu_1) | \psi_E
\rangle
\end{equation}
from which it follows that $\mathcal{R}_1| \psi_E \rangle \neq 0$
since $E \neq \mu_1 $. Applying the operator $\mathcal{R}_1$ to both
sides of Eq.(13), one obtains
\begin{equation}
\mathcal{R}_1 \mathcal{Q}_1 | \tilde{\psi}_E \rangle=(E-\mu_1) |
\tilde{\psi}_E \rangle,
\end{equation}
i.e. $\mathcal{H}_2 | \tilde{\psi}_E \rangle= E | \tilde{\psi}_E
\rangle$, where we have set $ | \tilde{\psi}_E \rangle=\mathcal{R}_1
|\psi_E \rangle$ or, explicitly [see Eq.(5)]
\begin{equation}
\tilde{\psi}_n(E)=r_n^{(1)} \psi_n(E)+\bar{r}_n^{(1)} \psi_{n-1}(E).
\end{equation}
Therefore, $ | \tilde{\psi}_E \rangle$ is an eigenfunction of
$\mathcal{H}_2$ corresponding to the energy $E$. Also, from Eqs.(6),
(7), (15) and from the assumed asymptotic behavior of
$\kappa_n^{(1)}$ and $V_n^{(1)}$ as $ n \rightarrow \pm \infty$, it
follows that $|\tilde{\psi}_E \rangle$ is a proper (improper)
eigenfunction of $\mathcal{H}_2$ in the same way as $|\psi_E
\rangle$ is a proper (improper) eigenfunction of $\mathcal{H}_1$. In
a similar way, one can show that any eigenvalue $E$ of
$\mathcal{H}_2$, belonging to its continuous or to its point
spectrum, is also an eigenvalue of $\mathcal{H}_1$ provided that $E
\neq \mu_1$. Therefore the continuous and point spectra of
$\mathcal{H}_1$ and $\mathcal{H}_2$ do coincide, apart from the
energy level $E=\mu_1$ which needs a separate analysis. For
$E=\mu_1$, the eigenvalue equation $\mathcal{H}_2 |\psi \rangle =
\mu_1 | \psi \rangle$ can be satisfied by taking $\mathcal{Q}_1 |
\psi \rangle =0$, which reads explicitly
\begin{equation}
q_n^{(1)} \psi_n+{q}_n^{(1)} \psi_{n+1}=0.
\end{equation}
Using the expressions of $q_n^{(1)}$ and $\bar{q}_n^{(1)}$ given by
Eqs.(6-9), the difference equation (16) for $\psi_n$ can be solved
in a closed form, yielding
\begin{equation}
\psi_n=\frac{1}{\sqrt{\kappa_n^{(1)} \phi^{(1)}_n \phi^{(1)}_{n-1}
}}.
\end{equation}
In view of the asymptotic behaviors of $\phi_n^{(1)}$ and $\kappa_n$
as $n \rightarrow \pm \infty$ and assuming that $\phi_n^{(1)}$ does
not vanish for any integer $n$, it turns out that $\psi_n$ is
bounded and $\psi_n \rightarrow 0 $ as $ n \rightarrow \pm \infty$,
i.e. $E=\mu_1$ belongs to the point spectrum of $\mathcal{H}_2$ and
its eigenfunction is given by Eq.(17).\\
It should be noted that the synthesis of the partner Hamiltonian
$\mathcal{H}_2$, with spectrum $\sigma_2=\sigma_1 \cup \{ \mu_1 \}$,
is not unique because of some freedom left in the choice of
$\phi_n^{(1)}$ satisfying Eq.(2) once $\mu_1$ has been fixed:
different choices of $\phi_n^{(1)}$ lead in fact to different
lattice realizations of $\mathcal{H}_2$, i.e. different values of
hopping amplitudes $\kappa_n^{(2)}$ and site energies $V_n^{(2)}$.
\par The factorization method
can be iterated to synthesize new Hamiltonians $\mathcal{H}_3$,
$\mathcal{H}_4$, $\mathcal{H}_5$, ... whose energy spectra differ
from that of $\mathcal{H}_1$ owing to the addition of the discrete
energy levels $\{ \mu_1, \mu_2 \}$, $\{ \mu_1, \mu_2, \mu_3 \}$, $\{
\mu_1, \mu_2, \mu_3, \mu_4  \}$, ..., with $|\mu_k|> 2 \kappa$ ($k=1,2,3,4,...$). \\
An interesting property, that is proven in the Appendix A, is the
following one. Let us assume $V_{n}^{(1)}=0$ for the lattice
Hamiltonian $\mathcal{H}_1$. Then a partner Hamiltonian
$\mathcal{H}_{2N+1}$, obtained from $\mathcal{H}_1$ by adding $2N$
new energy levels $\mu_1, \mu_2, \mu_3,...., \mu_{2N}$ with
$\mu_2=-\mu_1$, $\mu_4=-\mu_3$,...., $\mu_{2N}=-\mu_{2N-1}$, can be
synthesized in such a way that $V^{(2N+1)}_n=0$. This means that the
partner lattice described by $\mathcal{H}_{2N+1}$ and supporting
$2N$ bound states differs from the original one, defined by
$\mathcal{H}_1$, because of different hopping rates $\kappa_n$
between adjacent sites, but not for the site energies $V_n$.\\
As a final note, it should be mentioned that the technique of
intertwining operators so far described could generate non-Hermitian
lattice Hamiltonians with complex-valued hopping rates $\kappa_n$ or
site energies $V_n$, even though the initial Hamiltonian
$\mathcal{H}_1$ is Hermitian. However, in spite of non-Hermiticity,
the energy spectrum of such synthesized Hamiltonians remains by
construction real-valued. This situation is especially interesting
for the synthesis of invisible defects in the lattice, as discussed
in Sec.IV.

\section{Tight binding lattices with reflectionless defects}
The intertwining operator technique presented in the previous
section can be applied to the synthesis of lattices with
reflectionless defects. Previous works have so far limited to
consider Hermitian lattices (see, for instance,
\cite{dis1,Sukhorukov}); conversely, here we do not necessarily
require that the partner Hamiltonians $\mathcal{H}_2$,
$\mathcal{H}_3$, $\mathcal{H}_4$, ..., obtained by the iterated
application of intertwining operator method, be self-adjoint.
Notably, it will be shown in the next section that a truly
invisibility of the defects requires the synthesis of non-Hermitian
lattices. In this section, we first discuss the scattering
properties of partner lattice Hamiltonians obtained by the
intertwining operator technique, and then apply the results to the
synthesis of reflectionless defects in the lattices.

\subsection{Scattering properties of partner lattice Hamiltonians}
Let $\mathcal{H}_1$ and $\mathcal{H}_2$ be the Hamiltonians of the
two partner tight-binding lattices defined by Eqs.(3) and (10). By
construction, the two Hamiltonians have the same energy spectrum,
except for an additional energy level $\mu_1$ for $\mathcal{H}_2$.
The two lattices are homogeneous (i.e., free of defects) at $n
\rightarrow \pm \infty$; therefore, asymptotically they admit of
plane-wave solutions of the form $\sim \exp( \pm i q n)$, where $q$
is the wave number that varies in the interval $ 0 \leq q < \pi$.
Such plane waves belong to the common continuous spectrum of the
Hamiltonians, with energy $E(q)=2 \kappa \cos(q)$. The reflection
($r_1(q)$, $r_2(q)$) and transmission ($t_1(q)$, $t_2(q)$)
coefficients of the two lattices are defined by the asymptotic
behavior of scattered waves at $n \rightarrow \pm \infty$ from a
forward-incident plane wave $\sim \exp(-iqn)$ according to the
relations \cite{notegroup}
\begin{equation}
\psi_n^{(1)} \sim \left\{
\begin{array}{cc}
\exp(-iqn)+r_1(q) \exp(iqn) & n \rightarrow -\infty \\
t_1(q) \exp(-iqn) & n \rightarrow \infty
\end{array}
\right.
\end{equation}
for $\mathcal{H}_1$, and
\begin{equation}
\psi_n^{(2)} \sim \left\{
\begin{array}{cc}
\exp(-iqn)+r_2(q) \exp(iqn) & n \rightarrow -\infty \\
t_2(q) \exp(-iqn) & n \rightarrow \infty
\end{array}
\right.
\end{equation}
for $\mathcal{H}_2$. Let us indicate by $\omega_1$ the real-valued
and positive solution to the equation
\begin{equation}
|\mu_1|=2 \kappa \cosh(\omega_1)
\end{equation}
and let $\delta_1=\mu_1/|\mu_1|$ (i.e. $\delta_1=1$ for $\mu_1>0$,
$\delta_1=-1$ for $\mu_1<0$). It can be then proven that the
following relations between transmission and reflection coefficients
of the two partner Hamiltonians hold
\begin{eqnarray}
t_2(q)  =  t_1(q) \frac{\exp(-\omega_1/2)-\delta_1 \exp(
\omega_1/2+iq)}{\exp(\omega_1/2)-\delta_1 \exp( -\omega_1/2+iq)} \\
r_2(q)  =  r_1(q) \frac{\exp(\omega_1/2)-\delta_1 \exp(
-\omega_1/2-iq)}{\exp(\omega_1/2)-\delta_1 \exp( -\omega_1/2+iq)}.
\end{eqnarray}
The proof of Eqs.(21) and (22) is given in the Appendix B. Here we
just noticed that $|r^{(1)}(q)|=|r^{(2)}(q)|$ and
$|t^{(1)}(q)|=|t^{(2)}(q)|$, i.e. the transmittance and reflectance
coefficients of the two partner lattices are the same. It should be
noted that, as  $|t^{(1)}(q)|^2+|r^{(1)}(q)|^2=1$ for the Hermitian
$\mathcal{H}_1$  lattice, it follows that
$|t^{(2)}(q)|^2+|r^{(2)}(q)|^2=1$ either, even if the partner
Hamiltonian $\mathcal{H}_2$ is non-Hermitian. This result is a
non-trivial one because it is known that unitarity of the scattering
matrix in a generic non-Hermitian Hamiltonian  is usually broken,
and the reflection and transmission coefficients can be unbounded
(see, for instance, \cite{Mostafazadeh} and references therein).\\
By simple iteration, Eqs.(21) and (22) can be readily extended to
the case of the partner Hamiltonian $\mathcal{H}_N$ obtained from
$\mathcal{H}_1$ by adding the energy levels $\mu_1$, $\mu_2$, ...,
$\mu_N$. The reflection ($r_N(q)$) and transmission ($t_N(q)$)
coefficients of the lattice described by $\mathcal{H}_N$ are given
by
\begin{eqnarray}
t_N(q)  =  t_1(q) \prod_{k=1}^{N} \frac{\exp(-\omega_k/2)-\delta_k
\exp(
\omega_k/2+iq)}{\exp(\omega_k/2)-\delta_k \exp( -\omega_k/2+iq)} \; \; \\
r_N(q)  =  r_1(q) \prod_{k=1}^N \frac{\exp(\omega_k/2)-\delta_k
\exp( -\omega_k/2-iq)}{\exp(\omega_k/2)-\delta_k \exp(
-\omega_k/2+iq)} \; \;
\end{eqnarray}
where $\omega_k$ is the positive root of the equation $2 \kappa
\cosh(\omega_k)=|\mu_k|$ and $\delta_k=\mu_k/|\mu_k|$
($k=1,2,3,...,N$).

\subsection{Lattice with reflectionless defects}
 Reflectionless lattices containing
localized defects are readily synthesized by assuming for
$\mathcal{H}_1$ the Hamiltonian of a homogeneous and defect-free
lattice ($\kappa_n^{(1)}=1$, $V_n^{(1)}=0$), for which $r_1(q)=0$
and $t_1(q)=1$. In fact, from Eq.(24) it follows that the reflection
coefficient $r_N(q)$ of any partner Hamiltonian $\mathcal{H}_N$
vanishes, and the incident wave is fully transmitted through the
lattice. Depending on the choice of the sequences $\phi_n^{(1)}$,
$\phi_n^{(2)}$,  $\phi_n^{(3)}$, ..., the resulting partner
Hamiltonian may be or may not be Hermitian.\\
\\
{\em Hermitian Lattices}\\
\\
\begin{figure}[htbp]
  \includegraphics[width=85mm]{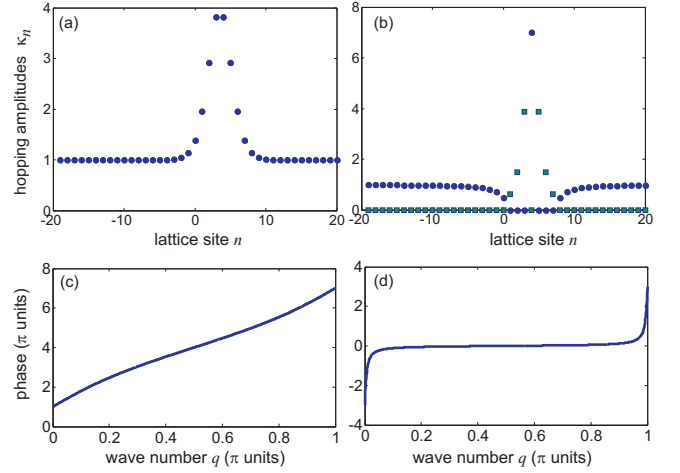}\\
   \caption{(color online) (a) Behavior of the hopping rates
$\kappa_n$ for a Hermitian lattice as predicted by Eq.(30) for
parameter values $N=3$, $\omega_1=0.6$ and $\alpha=0$. (b) Behavior
of the hopping rates $\kappa_n$ for a non-Hermitian lattice as
predicted by Eq.(35) for parameter values $N=3$, $\omega_1=0.01$ and
$\alpha=0.5$. In the figure, the dots refer to ${\rm Re}(\kappa_n)$,
whereas the squares to ${\rm Im}(\kappa_n)$. In (c) and (d) the
behaviors of the phase of the transmission coefficient $t(q)$ of the
two lattices are also depicted.}
\end{figure}
 Examples of reflectionless and Hermitian lattices obtained
by the application of the intertwining operator technique or by
other techniques have been previously presented in
\cite{dis1,dis2,Sukhorukov}. The simplest case corresponds to the
addition of a single energy level $\mu_1$ outside the tight-binding
band $-\kappa < E < \kappa$. Assuming for instance $\mu_1>\kappa$,
Eq.(2) can be satisfied with the choice
\begin{equation}
\phi_n^{(1)}= \cosh[\omega_1 (n-\alpha)]
\end{equation}
 which ensures the Hermiticity of the partner Hamiltonian
 $\mathcal{H}_2$. In Eq.(25), $\omega_1={\rm acosh (\mu_1/ 2 \kappa)}$ and $\alpha$ is an arbitrary real parameter. The hopping amplitudes
and site energies of the partner lattice read explicitly [see
Eqs.(11) and (12)]
\begin{eqnarray}
\kappa_n^{(2)} & = & \frac{\sqrt{\cosh[\omega_1(n-\alpha-2)]
\cosh[\omega_1(n-\alpha)]}}{\cosh[\omega_1(n-\alpha-1]} \\
 V_n^{(2)} & = & \frac{\cosh [\omega_1(n-\alpha+1)]}{\cosh [\omega_1(n-\alpha)]}-\frac{\cosh [\omega_1(n-\alpha)]}{\cosh [\omega_1
 (n-\alpha-1)]}. \;\;\;\;\;\;\;\;
\end{eqnarray}
Such a lattice, in spite of the presence of defects, is
reflectionless and supports one bound state, given by [see Eq.(17)]
\begin{equation}
\psi_n=\frac{1}{\sqrt{\cosh[\omega_1(n-\alpha)]
\cosh[\omega_1(n-\alpha-1)]}}.
\end{equation}
Another example, which was recently proposed in
Ref.\cite{Sukhorukov}, is provided by the partner lattice
$\mathcal{H}_3$ obtained from the defect-free lattice
$\mathcal{H}_1$ by adding the couple of energy levels $\mu_1>
\kappa$ and $\mu_2=-\mu_1$ \cite{note3}. In this case, assuming
again for $\phi_n^{(1)}$  the expression given by Eq.(25), according
to the analysis of Sec.II and Appendix A the hopping amplitudes of
the Hermitian lattice $\mathcal{H}_3$ read explicitly [see Eq.(A6)]
\begin{equation}
\kappa_n^{(3)}  =  \sqrt{\frac{\cosh[\omega_1(n-\alpha)]
\cosh[\omega_1(n-\alpha-3)]}{\cosh[\omega_1(n-\alpha-1)]
\cosh[\omega_1(n-\alpha-2)]}}
\end{equation}
whereas $V_n^{(3)} = 0$ for the site energies.
 The lattice $\mathcal{H}_3$ is, by construction,
reflectionless and supports two bound states. With the procedure
outlined in the previous section, Hermitian lattices supporting an
arbitrarily large number of bound states can be constructed in this
way. A simple and noteworthy case, which generalizes the previous
example, is provided by the lattice Hamiltonian $\mathcal{H}_{2N+1}$
obtained from the defect-free lattice $\mathcal{H}_1$ by adding the
$2N$ energy levels $\mu_1=2 \kappa \cosh(\omega_1)$, $\mu_2=-\mu_1$,
$\mu_3=2 \kappa \cosh( 2\omega_1)$, $\mu_4=-\mu_3$, ....,
$\mu_{2N-1}=2 \kappa \cosh(N \omega_1)$, $\mu_{2N}=-\mu_{2N-1}$. In
this case, with the choice (25) for $\phi_n^{(1)}$, one can show
that the hopping rates of the lattice $\mathcal{H}_{2N+1}$ take the
simple form \cite{dis1}
\begin{equation}
\kappa_n^{(2N+1)}  =  \sqrt{\frac{\cosh[\omega_1(n-\alpha)]
\cosh[\omega_1(n-\alpha-2N-1)]}{\cosh[\omega_1(n-\alpha-N)]
\cosh[\omega_1(n-\alpha-N-1)]}}
\end{equation}
which generalizes Eq.(29). An an example, Fig.1(a) shows the
behavior of the hopping rates $k_n$, as predicted by Eq.(30), for
the case $N=3$ and for $\omega_1=0.6$, $\alpha=0$. As shown in the
next section, even though being reflectionless, such Hermitian
lattices are not invisible owing to the energy-dependence introduced
by the bound
states in the phase of the transmission coefficient.\\
\\
{\em Non-Hermitian Lattices}\\
\\
A different choice of the sequences $\phi_n^{(1)}$, $\phi_n^{(2)}$,
... can be used to synthesized reflectionless non-Hermitian
lattices. The simplest case corresponds, as in the previous
Hermitian case, to the addition of a single energy level $\mu_1$
outside the tight-binding band $-\kappa < E < \kappa$.\\
Let us assume, for the sake of definiteness, $\mu_1>\kappa$ and let
us make the choice [which replaces Eq.(25)]
\begin{equation}
\phi_n^{(1)}= \sinh[\omega_1 (n-\alpha)],
\end{equation}
 where $\omega_1={\rm acosh (\mu_1/ 2 \kappa)}$ and $\alpha$ is an arbitrary real (but non-integer)
 parameter. The expressions of hopping amplitudes and site energies of the partner lattice
Hamiltonian $\mathcal{H}_2$ are then given by
\begin{eqnarray}
\kappa_n^{(2)} & = & \sqrt{ \frac{\sinh[\omega_1(n-\alpha-2)]
\sinh[\omega_1(n-\alpha)]}{\sinh^2[\omega_1(n-\alpha-1)]}} \\
 V_n^{(2)} & = & \frac{\sinh[\omega_1(n-\alpha+1)]}{\sinh [\omega_1(n-\alpha)]}-\frac{\sinh [\omega_1(n-\alpha)]}{\sinh [\omega_1
 (n-\alpha-1)]}. \;\;\;\;\;\;\;\;
\end{eqnarray}
which replace Eqs.(26) and (27), respectively. Note that, as the
site energies $V_n^{(2)}$ are always real-valued, the hopping
amplitudes $\kappa_n^{(2)}$ are not. Specifically, $\kappa_n^{(2)}$
becomes purely imaginary at the two lattice sites $n$ satisfying the
condition $ \alpha < n < 2+\alpha$. Therefore, the partner
Hamiltonian $\mathcal{H}_2$ is not Hermitian, in spite its spectrum
is real-valued by construction.\\
As a second example, let us synthesize the partner lattice
$\mathcal{H}_3$ obtained from the defect-free lattice
$\mathcal{H}_1$ by adding the couple of energy levels $\mu_1>
\kappa$ and $\mu_2=-\mu_1$, assuming again for $\phi_n^{(1)}$  the
expression given by Eq.(31).  According to the analysis of Sec.II
and Appendix A, the hopping amplitudes of the lattice
$\mathcal{H}_3$ now read explicitly [compare with Eq.(29)]
\begin{equation}
\kappa_n^{(3)} = \sqrt{\frac{\sinh[\omega_1(n-\alpha)]
\sinh[\omega_1(n-\alpha-3)]}{\sinh[\omega_1(n-\alpha-1)]
\sinh[\omega_1(n-\alpha-2)]}}
\end{equation}
whereas $V_n^{(3)}=0$ for the site energies. By construction, the
lattice Hamiltonian $\mathcal{H}_3$ is reflectionless, has a
real-valued energy spectrum and supports two bound states,
corresponding to the energies $E=\pm 2 \kappa \cosh(\omega_1)$.
However, an inspection of Eq.(34) indicates that $\mathcal{H}_3$ is
not Hermitian because the hopping amplitudes $\kappa_n^{(3)}$ take
an imaginary value at the two sites $n$ satisfying the condition
$\alpha < n < 1+\alpha$ and $2+ \alpha < n < 3+\alpha$. More
generally, with the choice (31), the  non-Hermitian Hamiltonian
$\mathcal{H}_{2N+1}$ admitting $2N$ bound states with energies
$\mu_1=2 \kappa \cosh(\omega_1)$, $\mu_2=-\mu_1$, $\mu_3=2 \kappa
\cosh( 2\omega_1)$, $\mu_4=-\mu_3$, ...., $\mu_{2N-1}=2 \kappa
\cosh(N \omega_1)$, $\mu_{2N}=-\mu_{2N-1}$ can be synthesized,
corresponding to the hopping amplitudes [compare with Eq.(30)]
\begin{equation}
\kappa_n^{(2N+1)}  =  \sqrt{\frac{\sinh[\omega_1(n-\alpha)]
\sinh[\omega_1(n-\alpha-2N-1)]}{\sinh[\omega_1(n-\alpha-N)]
\sinh[\omega_1(n-\alpha-N-1)]}}
\end{equation}
and site energies $V_n^{(2N+1)}=0$. Note that the hopping amplitudes
are purely imaginary at lattice sites $n$ satisfying the conditions
$\alpha < n <\alpha+N$ and $\alpha+N+1 < n <\alpha+2N+1$. As an
example, Fig.1(b) shows the behavior of the real and imaginary parts
 of the hopping amplitudes $\kappa_n$ as given by Eq.(35) for
$N=3$, $\omega=0.01$ and $\alpha=0.5$.\\
\begin{figure}[htbp]
  \includegraphics[width=65mm]{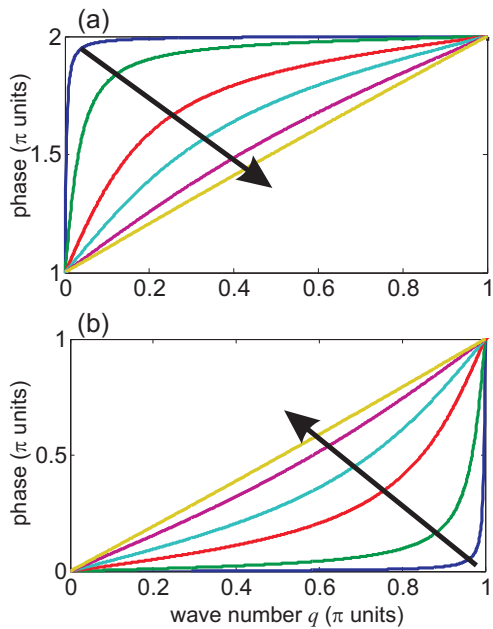}\\
   \caption{(color online) Behavior of the phase $\varphi_k(q)$ versus the wave number $q$ (in units of
   $\pi$),
   defined by Eq.(37), for (a) $\delta_k=1$, (b) $\delta_k=-1$ and for
   increasing values of $\omega_k$. The curves refer to the values $\omega_k=0.01$, $0.1$, $0.5$, $1$, $2$ and $4$.
   The arrows in the figures show the direction of increasing
   $\omega_k$.}
\end{figure}
\begin{figure}[htbp]
  \includegraphics[width=80mm]{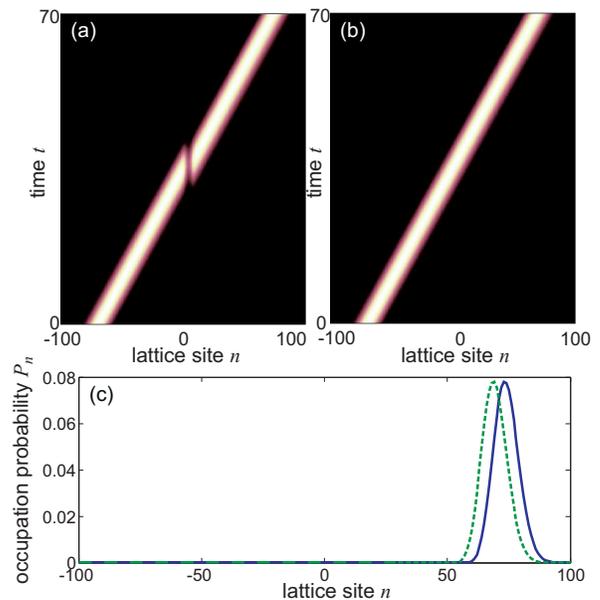}\\
   \caption{(color online) Propagation of an initial Gaussian-shaped wave
packet (snapshot of the site occupation probabilities
$P_n(t)=|\langle \psi(t)|n \rangle|^2$) (a) in the Hermitian lattice
with hopping rates shown in Fig.1(a) [parameter values are given in
the text], and (b) in the defect-free lattice. In (c) the behaviors
of site occupation probabilities $P_n(t)$ at time $t=70$ in the two
lattices are depicted (the solid line refers to the Hermitian
lattice with defects, the dashed line to the defect-free lattice).
Note the advancement experienced by the wave packet propagating in
the lattice with defects. Such an advancement is basically
ascribable to the increase of hopping rates $\kappa_n$ in the defect
region [see Fig.1(a)].}
\end{figure}
One could wonder whether non-Hermitian tight-binding lattices with
imaginary hopping amplitudes may describe wave transport in some
physically realizable systems. Coupled optical waveguide structures
with gain and/or loss regions have been recently proposed as
experimentally accessible systems to mimic the dynamics of
non-Hermitian lattices with complex-valued site energies (see, for
instance, \cite{TB3,TB4,TB5,Christodoulides}); however, the
non-Hermitian lattices discussed in the previous examples require
imaginary values of the hopping rates at some site energies, an
issue which was not considered in such previous works. In the
Appendix C, it is shown that suitable {\em longitudinal} modulations
of gain/loss and propagation constants in evanescently-coupled
optical waveguide arrays lead to an effective non-Hermitian lattice
with imaginary hopping amplitudes that realizes the models discussed
in this section.

\section{Invisibility in non-Hermitian lattices}

For a reflectionless lattice synthesized by the intertwining
operator technique, the transmission coefficient as a function of
the wave number $q$ of the incident wave has the form $t(q)= \exp [i
\varphi(q)]$, where according to Eq.(23) the phase $\varphi(q)$ is
given by the sum of $N$ contributions associated to each of the $N$
bound states with energies $\mu_1$, $\mu_2$, ..., $\mu_N$, i.e.
\begin{equation}
\varphi(q)=\sum_{k=1}^{N} \varphi_k(q)
\end{equation}
where
\begin{equation}
\exp[i \varphi_k(q)]=\frac{\exp(-\omega_k/2)-\delta_k \exp(
\omega_k/2+iq)}{\exp(\omega_k/2)-\delta_k \exp( -\omega_k/2+iq)},
\end{equation}
\begin{figure}[htbp]
  \includegraphics[width=80mm]{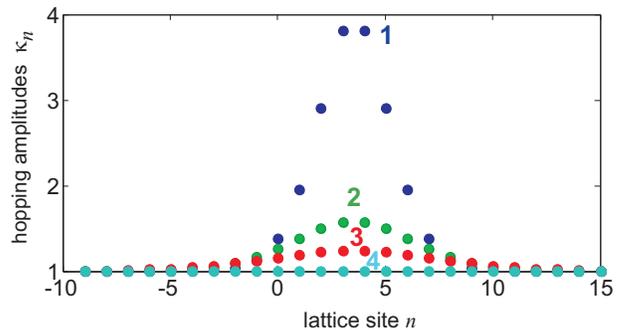}\\
   \caption{(color online) Behavior of the hopping rates $\kappa_n$ for the
   Hermitian lattice, as given by Eq.(30), for $N=3$, $\alpha=0$ and for decreasing values of $\omega_1$ (curve 1:
   $\omega_1=0.6$;
   curve 2: $\omega_1=0.3$; curve 3: $\omega_1=0.2$; curve 4: $\omega_1=0.02$).}
\end{figure}
$\mu_k=2 \kappa \delta_k \cosh(\omega_k)$, $\delta_k=\mu_k/|\mu_k|$,
and $\omega_k>0$ ($k=1,2,3,...,N$). The behavior of $\varphi_k(q)$,
for increasing values of $\omega_k$ and for $\delta_k= \pm 1$, is
shown in Fig.2. In case $\delta_k=1$ [i.e. $\mu_k>0$, see Fig.2(a)],
one has $\varphi_k(q) \simeq \pi + q$ for $\omega_k \gg 1$ and
$\varphi_k(q) \rightarrow 0$ ($\rm{mod} \; 2 \pi$, $q \neq 0$) for
$\omega_k \rightarrow 0^+$. Similarly, in case $\delta_k=-1$ [i.e.
$\mu_k<0$, see Fig.2(b)], one has $\varphi_k(q) \simeq q$ for
$\omega_k \gg 1$ and $\varphi_k(q) \rightarrow 0$ ($q \neq \pi$) for
$\omega_k \rightarrow 0^+$. Note that, according to Eq.(36), the
behavior of the overall phase $\varphi(q)$ is given by the
superposition of the various terms $\varphi_k(q)$ and does not
depend on whether the synthesized partner
Hamiltonian $\mathcal{H}_N$ is Hermitian or non-Hermitian. \\
We now ask ourselves whether the defects in the partner lattice, in
addition of being reflectionless,  are also {\em invisible} to an
outside observer. This condition requires that the phase
$\varphi(q)$ of the transmission coefficient be flat, i.e. that $(d
\varphi/dq)=0$ almost everywhere. If this condition is not
satisfied, the spectral components of a wave packet crossing the
defect region of the partner lattice would acquire the additional
phase contribution $\varphi(q)$, absent in the defect-free lattice,
which would be responsible for a {\em different} time-of-flight and
for a {\em different} distortion of the wave packet as compared to
the same wave packet propagating in the ideal defect-free lattice.
Therefore, an outside observer could detect the existence of defects
somewhere in the lattice by e.g. simple time-of-flight measurements.
The {\em advance} in the time of flight experienced by the wave
packet propagating in the partner lattice with defects can be
readily calculated by standard methods of phase or group-delay time
analysis, and reads
\begin{equation}
\tau_g=\frac{1}{v_g} \left( \frac{d \varphi}{d q}
\right)_{q_0}=\frac{1}{2 \kappa \sin (q_0)} \left(
\frac{d\varphi}{dq} \right)_{q_0},
\end{equation}
\begin{figure}[htbp]
  \includegraphics[width=80mm]{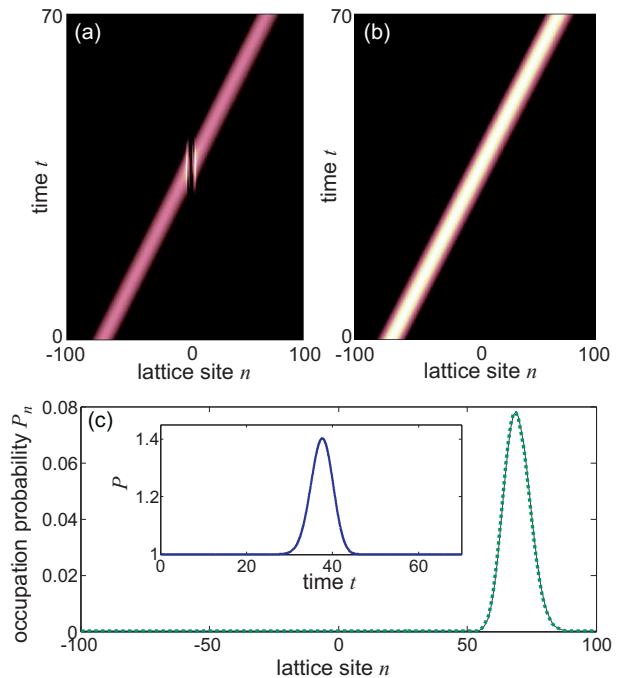}\\
   \caption{(color online) Propagation of an initial Gaussian-shaped wave
packet (snapshot of the site occupation probabilities
$P_n(t)=|\langle \psi(t)|n \rangle|^2$) (a) in the non-Hermitian
lattice with hopping rates shown in Fig.1(b) [parameter values are
given in the text], and (b) in the defect-free lattice. In (c) the
behaviors of site occupation probabilities $P_n(t)$ at time $t=70$
in the two lattices are depicted (the thin solid line refers to the
non-Hermitian lattice with defects; the dashed line, almost
overlapped with the solid one, refers to the defect-free lattice).
The inset in (c) shows the behavior of the total occupation
probability $P(t)=\sum_n P_n(t)$ versus time in the non-Hermitian
lattice of (a).}
\end{figure}
where $q_0$ is the carrier wave number of the wave packet and $v_g=2
\kappa \sin(q_0)>0$ its group velocity. In particular, for a partner
lattice synthesized by taking $\omega_k \gg 1$, one has $(d
\varphi(q)/dq) \simeq N$ (see Fig.2), and thus the advancement of
the wave packet measured by an outside observer (i.e. far from the
defect region) would be $\sim N/v_g$. Hence, comparing the time of
flight measurements in the two lattices, the observer can estimate
the number $N$ of bound states of the partner lattice. From the
above considerations, it follows that the necessary and sufficient
condition for a reflectionless lattice to be also invisible is that
$\omega_k \rightarrow 0$. For a Hermitian lattice, from Eqs.(26) and
(27) [and similarly from Eqs.(29) or (30)] it follows that in this
limit the lattice becomes defect-free, i.e. $\kappa_n \rightarrow 1$
and $V_n \rightarrow 0$ regardless of the value of the parameter
$\alpha$. This means that, for the Hermitian lattices synthesized in
Sec.III.B, the invisibility condition  is the absence of defects.
Conversely, from Eqs.(32) and (33) [and similarly from Eqs.(34) or
(35)] it follows that, in the $\omega_k \rightarrow 0$ limit,
$\kappa_n$ and $V_n$ do not tend to the values of the defect-free
lattice [see, for instance, Fig.1(b)]. This means that, in the
non-Hermitian lattices synthesized in Sec.III.B, invisibility of
defects can be achieved. It should be noted that such non-Hermitian
lattices with localized defects possessing a real-valued energy
spectrum are not $\mathcal{PT}$ invariant, i.e. $\mathcal{PT}$
symmetry is not of relevance for the achievement of
invisibility of the defects. \\
We have checked these predictions by direct numerical simulations of
wave packet propagation in Hermitian and non-Hermitian tight-binding
lattices with zero site energies and with hopping amplitudes defined
according to Eqs.(30) and (35), respectively. As an example,
Fig.3(a) shows the propagation of an initial Gaussian-shaped wave
packet $|\psi(t=0) \rangle= \sum_n \mathcal{N }\exp[-(n+n_0)^2/w^2]
\exp(-iq_0 n) |n \rangle$ in an Hermitian lattice with hopping rates
given by Eq.(30) for parameter values $\kappa=1$, $N=3$,
$\omega_1=0.6$, $\alpha=0$, $n_0=70$, $w_0=10$ and $q_0=\pi/2$
($\mathcal{N}$ is the normalization constant). The profile of
hopping rates for this lattice was shown in Fig.1(a). For
comparison, Fig.3(b) shows the propagation of the same wave packet
in the defect-free lattice. The distribution of site occupation
probabilities $P_n(t)=|\langle n | \psi(t) \rangle|^2$ at time
$t=70$ in the two cases is shown in Fig.3(c). Note that, according
to the previous analysis, the wave packet is fully transmitted in
both lattices, and far from the inhomogeneities it propagates with
the group velocity $v_g=2 \kappa \sin(q_0)=2$. However, in the
lattice with defects the wave packet is advanced, as one can see
clearly from an inspection of Fig.3(c). The behavior of the phase
$\varphi(q)$ of the transmission coefficient of the partner lattice
corresponding to the simulation of Fig.3(a) is shown in Fig.1(c).
One might think that, to make the Hermitian lattice invisible, one
should reduce the value of $\omega_1$; however, as discussed
previously and as shown in Fig.4, as  $\omega_1$ is diminished
toward zero, the defects in the hopping amplitudes vanish and the
lattice basically becomes defect-free. Conversely, Fig.5 shows that
a non-Hermitian lattice can be invisible yet presenting defects in
the hopping amplitudes. Figure 5(a) shows the propagation of the
same initial Gaussian-shaped wave packet $|\psi(t=0) \rangle= \sum_n
\mathcal{N} \exp[-(n+n_0)^2/w^2] \exp(-iq_0 n) | n \rangle$ of
Fig.3, but in the non-Hermitian lattice with hopping rates given by
Eq.(35) for parameter values $\kappa=1$, $N=3$, $\omega_1=0.01$ and
$\alpha=0.5$ [the distribution of hopping rates for this lattice was
shown in Fig.1(b)]. For comparison, Fig.5(b) shows the propagation
of the same wave packet in the defect-free lattice. The distribution
of site occupation probabilities $P_n(t)=|\langle n | \psi(t)
\rangle|^2$ at time $t=70$ in the two cases is shown in Fig.5(c).
Note that, owing to the flatness of the phase $\varphi(q)$ for this
lattice [see Fig.1(d)], the wave packet is fully transmitted with no
appreciable delay and/or distortion, as one can infer from an
inspection of Fig.5(c). An outside observer thus can not distinguish
whether the transmitted wave packet has been propagated in a
defect-free or in an inhomogeneous lattice, and thus the defects in
the non-Hermitian lattice are fully invisible. It should be finally
noted that the total probability $P(t)=\sum_n P_n(t)$ is transiently
not conserved in the non-Hermitian lattice, and turns out to be
amplified during interaction with defects, as shown in the inset of
Fig.5(c). Such an enhancement of the probability, however, is not
visible to the outside observer.

\section{Conclusions}
In this work we have investigated theoretically the issue of
invisibility of reflectionless tight-binding lattices with defects
synthesized by the intertwining operator technique of supersymmetric
quantum mechanics. As for Hermitian lattices the defects are not
invisible and time-of-flight measurements of wave packets crossing
the defect region may reveal their existence, in this work it has
been shown that, in a certain class of non-Hermitian lattices with
complex hopping amplitudes, the defects may appear fully invisible
to an outside observer. In spite of non-Hermiticity, such lattices
have a real-valued energy spectrum. As discussed in the Appendix C,
arrays of evanescently-coupled optical waveguides with suitable
longitudinal modulation of loss/gain coefficients and propagation
constants could provide a physically realizable system to test
invisibility in non-Hermitian tight-binding lattices.

\acknowledgments
 This work was supported by the italian MIUR
(PRIN-2008 project "Analogie ottico-quantistiche in strutture
fotoniche a guida d'onda").

\appendix
\section{}
In this Appendix, the following theorem is proved:\\
Let $\mathcal{H}_1$ be a tight binding Hamiltonian  with
$V_{n}^{(1)}=0$, and let $\mathcal{H}_{2N+1}$ a partner Hamiltonian
synthesized from $\mathcal{H}_1$ by adding $2N$ new energy levels
$\mu_1, \mu_2, \mu_3,...., \mu_{2N}$, with $\mu_2=-\mu_1$,
$\mu_4=-\mu_3$,....,
$\mu_{2N}=-\mu_{2N-1}$. Then $\mathcal{H}_{2N+1}$ can be constructed in such a way that $V^{(2N+1)}_n=0$.\\
Let us first prove the theorem for $N=1$. The partner Hamiltonian
$\mathcal{H}_2$ is first constructed following the procedure
described in Sec.II, and the corresponding hopping amplitudes
$\kappa^{(2)}_n$ and site energies $V_n^{(2)}$ are given by Eqs.(11)
and (12), respectively, with $V_n^{(1)}=0$. To synthesize the
Hamiltonian $\mathcal{H}_3$, we need to construct the sequence
$\phi_n^{(2)}$ satisfying the difference equation
\begin{equation}
\kappa_n^{(2)} \phi^{(2)}_{n-1}+\kappa_{n+1}^{(2)}
\phi^{(2)}_{n+1}+V_{n}^{(2)} \phi_n^{(2)}=\mu_2 \phi^{(2)}_n
\end{equation}
and apply again the intertwining operator technique after the
factorization $\mathcal{H}_2=\mathcal{Q}_2 \mathcal{R}_2+\mu_2$. In
Eq.(A1), $\mu_2=-\mu_1$ and the asymptotic behavior $|\phi^{(2)}_n|
\rightarrow \infty$ for $n \rightarrow \pm \infty$ should be
satisfied. A possible choice for the sequence $\phi_n^{(2)}$ can be
obtained by observing that, since $V_n^{(1)}=0$, from Eq.(2) it
follows that $| \psi \rangle = \sum_n (-1)^n \phi^{(1)}_n |n
\rangle$ satisfies the equation $\mathcal{H}_1 |\psi \rangle =
-\mu_1 | \psi \rangle$, and thus $|\phi^{(2)} \rangle =
\mathcal{R}_1 | \psi \rangle$ satisfies the equation $\mathcal{H}_2
|\phi^{(2)} \rangle = -\mu_1 | \phi^{(2)} \rangle$, which is
precisely Eq.(A1). Using Eqs.(6), (7) and (15) one obtains after
some algebra
\begin{equation}
\phi^{(2)}_n=-2 (-1)^n \sqrt{\kappa_n^{(1)}
\phi_n^{(1)}\phi_{n-1}^{(1)}}.
\end{equation}
The hopping rates $\kappa_n^{(3)}$ and site energies of the partner
Hamiltonian $\mathcal{H}_3=\mathcal{R}_2\mathcal{Q}_2+\mu_2$,
obtained from $\mathcal{H}_2$ after changing the order of the
operators $\mathcal{Q}_2$ and $\mathcal{R}_2$, are then given by
[see Eqs.(11) and (12)]
\begin{eqnarray}
\kappa^{(3)}_n & = & \kappa_n^{(2)} \frac{r^{(2)}_{n-1}}{r^{(2)}_n} \\
V_n^{(3)} & = & V_n^{(2)}+\kappa_{n+1}^{(2)}
\frac{\phi_{n+1}^{(2)}}{\phi_n^{(2)}}-\kappa_n^{(2)}
\frac{\phi^{(2)}_n}{\phi^{(2)}_{n-1}}
\end{eqnarray}
where
\begin{equation}
r_n^{(2)} =  - \sqrt{\frac{\kappa_n^{(2)}
\phi_{n-1}^{(2)}}{\phi_n^{(2)}}}.
\end{equation}
Using in Eq. (A5) the expressions of $\kappa_n^{(2)}$ defined by
Eqs.(6) and (11), and of $\phi_n^{(2)}$ as given by Eq.(A2),
substitution of Eq.(A5) into Eqs.(A3) and (A4) finally yields after
some straightforward though lengthy algebra
\begin{eqnarray}
\kappa_n^{(3)} & = & \sqrt{ \frac{\kappa_n^{(1)} \kappa_{n-2}^{(1)}
\phi_n^{(1)} \phi_{n-3}^{(1)}}{\phi_{n-1}^{(1)}\phi_{n-2}^{(1)}} }\\
V_n^{(3)} & = & 0.
\end{eqnarray}
Therefore, for the partner Hamiltonian $\mathcal{H}_3$, obtained
from $\mathcal{H}_1$ by adding the two energies $\mu_1$ and
$\mu_2=-\mu_1$ with the procedure described above, one has
$V_n^{(3)}=0$. Starting from $\mathcal{H}_3$, one can repeat the
procedure to construct a partner Hamiltonian $\mathcal{H}_5$ with
  $V_n^{(5)}=0$ by adding to $\mathcal{H}_3$ the couple of eigenvalues $\mu_3$ and
$\mu_4=-\mu_3$. The hopping amplitudes $\kappa_n^{(5)} $of the new
Hamiltonian will be given by Eq.(A6), with $\kappa_n^{(1)}$ and
$\phi_n^{(1)}$ replaced by $\kappa_n^{(3)}$ and $\phi_n^{(3)}$,
respectively. By induction, it follows that a partner Hamiltonian
$\mathcal{H}_{2N+1}$, obtained from $\mathcal{H}_1$ by adding $N$
couples of energies $\{ \mu_1, \mu_2=-\mu_1\}$, $\{ \mu_3,
\mu_4=-\mu_3\}$, ..., $\{ \mu_{2N-1}, \mu_{2N}=-\mu_{2N-1}\}$, can
be always synthesized to have $V_n^{(2N+1)}=0$, which proves the
theorem.

\section {}
In this Appendix we prove the Eqs.(21) and (22) given in the text
relating the reflection and transmission coefficients of the two
partner lattice Hamiltonians $\mathcal{H}_1$ and $\mathcal{H}_2$. To
this aim, let us first consider the case $\mu_1>2 \kappa$, and let
us indicate by $\omega_1$ the positive root of the equation $\mu_1=
2 \kappa \cosh(\omega_1)$. As $\kappa_n^{(1)} \rightarrow \kappa$
and $V_n^{(1)} \rightarrow 0$ at $n \rightarrow \pm \infty$, the
asymptotic behavior of $\phi_n^{(1)}$, satisfying Eq.(2), is of the
form
\begin{equation}
\phi_n^{(1)} \sim \left\{
\begin{array}{cc}
\alpha \exp(\omega_1 n) & n \rightarrow + \infty \\
\beta \exp(-\omega_1 n) & n \rightarrow - \infty
\end{array}
 \right. ,
\end{equation}
where $\alpha$ and $\beta$ are two non-vanishing constants. From
Eqs.(6), (7), (11) and (12) it then follows that
\begin{eqnarray}
r_n^{(1)} & \rightarrow & -\exp( \mp \omega_1/2) \;\;\;\; {\rm  for}
\; n
\rightarrow \pm \infty \\
\bar{r}_n^{(1)} & \rightarrow & \exp( \pm \omega_1/2)  \;\;\;\; {\rm
for} \; n
\rightarrow \pm \infty \\
\kappa_n^{(2)} & \rightarrow & 1  \;\;\;\;  {\rm  for} \; n
\rightarrow \pm \infty \\
V_n^{(2)} & \rightarrow & 0   \;\;\;\; {\rm  for} \; n \rightarrow
\pm \infty.
\end{eqnarray}
Let us then indicate by $|\psi^{(1)} \rangle =\sum_n \psi_n^{(1)} |
n \rangle$ the solution to the equation $\mathcal{H}_1 | \psi^{(1)}
\rangle =E  | \psi^{(1)} \rangle$ corresponding to the scattering of
a forward propagating plane wave (coming from $n \rightarrow
-\infty$) with wave number $q$ and energy $E=2 \kappa \cos(q)$ ($0
\leq q < \pi$). The eigenfunction $| \psi^{(1)} \rangle$ has
therefore the asymptotic behavior expressed by Eq.(18) given in the
text. According to Eq.(15), the function $|\psi^{(2)} \rangle =
\sum_n \psi_n^{(2)} | n \rangle$ with
\begin{equation}
\psi_n^{(2)}=r_n^{(1)} \psi_n^{(1)}+\bar{r}_{n}^{(1)}
\psi_{n-1}^{(1)}
\end{equation}
satisfies the equation $\mathcal{H}_2 | \psi^{(2)} \rangle=E  |
\psi^{(2)} \rangle$. Using Eqs.(18), (B2) and (B3), it follows that
the asymptotic behavior of $\psi_n^{(2)}$ is given by
\begin{widetext}
\begin{equation}
\psi_n^{(2)} \sim \left\{
\begin{array}{cc}
\left[-\exp(\omega_1/2)+\exp(-\omega_1/2+iq) \right]
\exp(-iqn)+r_1(q) \left[\exp(-\omega_1/2-iq)-\exp(\omega_1/2)
\right] \exp(iqn) & n \rightarrow
-\infty \\
t_1(q) \left[-\exp(\omega_1/2)+\exp(\omega_1/2+iq) \right]
\exp(-iqn) & n \rightarrow \infty
\end{array}
\right.
\end{equation}
\end{widetext}
i.e. $|\psi^{(2)} \rangle$ describes the scattering, in the lattice
$\mathcal{H}_2$, of a plane wave with wave number $q$ coming from $n
\rightarrow -\infty$ and with amplitude
$[-\exp(\omega_1/2)+\exp(-\omega_1/2+iq)]$.  From Eq.(B7), the
transmission ($t_2$) and reflection ($r_2$) coefficients of the
partner lattice $\mathcal{H}_2$ are readily calculated, obtaining
the expressions (21) and (22) given in the text with $\delta_1=1$.\\
Let us now consider the case $\mu_1< -2 \kappa$, and let us indicate
again by $\omega_1$ the positive root of the equation $2 \kappa
\cosh(\omega_1)=-\mu_1$. The asymptotic behavior of $\phi_n^{(1)}$,
satisfying Eq.(2), is now of the form
\begin{equation}
\phi_n^{(1)} \sim \left\{
\begin{array}{cc}
\alpha (-1)^n \exp(\omega_1 n) & n \rightarrow + \infty \\
\beta (-1)^n \exp(-\omega_1 n) & n \rightarrow - \infty
\end{array}
 \right. ,
\end{equation}
where $\alpha$ and $\beta$ are again two non-vanishing constants. In
this case, the  asymptotic behavior of $r_n^{(1)}$ and
$\bar{r}_n^{(1)}$, as obtained from Eqs.(6), (7), and (B8), is given
by
\begin{eqnarray}
r_n^{(1)} & \rightarrow & -i \exp( \mp \omega_1/2) \;\;\;\; {\rm
for} \; n
\rightarrow \pm \infty \\
\bar{r}_n^{(1)} & \rightarrow & -i \exp( \pm \omega_1/2)  \;\;\;\;
{\rm for} \; n \rightarrow \pm \infty.
\end{eqnarray}
As compared to the previous case $\mu_1>0$, from Eqs.(B6), (B9) and
(B10) it follows that the asymptotic behavior of $\psi_n^{(2)}$ is
now given by the equation
\begin{widetext}
\begin{equation}
\psi_n^{(2)} \sim \left\{
\begin{array}{cc}
-i \left[\exp(\omega_1/2)+ \exp(-\omega_1/2+iq) \right] \exp(-iqn)-i
r_1(q) \left[ \exp(-\omega_1/2-iq)+ \exp(\omega_1/2) \right]
\exp(iqn) & n \rightarrow
-\infty \\
-i t_1(q) \left[\exp(-\omega_1/2)+\exp(\omega_1/2+iq) \right]
\exp(-iqn) & n \rightarrow \infty
\end{array}
\right.
\end{equation}
\end{widetext}
which replaces Eq.(B7). The transmission and reflection coefficients
$t_2$ and $r_2$ of the lattice $\mathcal{H}_2$ are readily
calculated from Eq.(B11), and their expressions are given by
Eqs.(21) and (22) with $\delta_1=-1$.

\section{}
In this Appendix we briefly discuss a possible physical realization
of non-Hermitian tight-binding lattices with complex hopping rates,
such as those discussed in Secs.III.B and IV. In the optical
context, it is known that Hermitian lattices can be implemented by
considering light propagation in arrays of evanescently-coupled
optical waveguides, the propagation direction $z$ of light playing
the role of time $t$ in the quantum-mechanical problem (see, for
instance, \cite{Sukhorukov,Longhi}). The evolution along $z$ of the
modal amplitudes $c_n$ of light trapped in the various waveguides of
the array is governed by the tight-binding Hamiltonian (1), in which
the site energies $V_n$ and hopping amplitudes $\kappa_n$ can be
engineered by a suitable design of waveguide channel widths, index
changes of the guiding cores, and distances between adjacent
waveguides in the array. In ordinary arrays, i.e. without loss or
gain regions, $V_n$ and $\kappa_n$ turn out to be real-valued, and
thus the Hamiltonian $\mathcal{H}$ Hermitian. Non-Hermitian lattices
with complex site energies can be mimicked by considering arrays of
evanescently-coupled waveguides in which light propagation in each
waveguide is either absorbed or amplified by some loss or gain
mechanism (see, for instance, \cite{TB5,Christodoulides}), where the
$z$-invariant gain or loss coefficients in the various waveguides
determine the imaginary parts of the site energies $V_n$. Such
non-Hermitian lattices have been intensively investigated in the
past few years, especially in connection with
$\mathcal{PT}$-symmetric quantum mechanics
\cite{TB3,TB4,TB5,Christodoulides}. However, the non-Hermitian
lattices that realize invisibility, discussed in Secs. III.B and IV,
have real-valued site energies $V_n$ but imaginary hopping rates
$\kappa_n$ at some lattice sites. To implement in optics such
invisible lattices, let us consider an array of evanescently-coupled
waveguides and assume that a suitable {\em longitudinal} and {\em
periodic} modulation of both gain/loss coefficient and effective
modal index, with spatial period $\Lambda$, is impressed to some
waveguides in the lattice. In this case, coupled-mode equations
describing the evolution of the modal amplitudes $c_n$ of light
trapped in the various waveguides read (see, for instance,
\cite{Longhi})
\begin{equation}
i \frac{dc_n}{dz}= \Delta_n c_{n-1}+\Delta_{n+1} c_{n+1}+ \left[
V_n+ \beta_n(z)-i \gamma_n(z) \right] c_n
\end{equation}
where $\Delta_n$ is the (real-valued) coupling rate between
waveguides $n$ and $n+1$, $V_n$ is the propagation constant mismatch
from a reference value, and $\beta_n(z)$, $\gamma_n(z)$ are the
impressed longitudinal modulations of the propagation constant and
loss/gain coefficient, respectively. We assume that both
$\beta_n(z)$ and $\gamma_n(z)$ are periodic functions, with spatial
period $\Lambda$ and with zero mean. This means that, on average, a
light field propagating in a {\em single} waveguide of the array
would not be damped nor amplified. Assuming that the spatial period
$\Lambda$ of the modulation is much shorter than the typical
coupling lengths ($\sim 1/ \Delta_n$) and mismatch lengths ($\sim
1/V_n$), after introduction of the amplitudes
\begin{equation}
a_n(z)=c_n(z) \exp \left[ i \varphi_n(z) \right]
\end{equation}
where
\begin{equation}
\varphi_n(z)=\int_0^z d \xi  \left[ \beta_n (\xi)-i\gamma_n (\xi)
\right],
\end{equation}
a set of effective equations for the slowly-varying amplitudes
$a_n(z)$ can be derived by a multiple-scale analysis (see, for
instance, \cite{Longhi08}). They read explicitly
\begin{eqnarray}
i \frac{d a_n}{dz} & = & \Delta_{n}  \langle \exp[i
\varphi_n(z)-i\varphi_{n-1}(z)] \rangle a_{n-1} + \\
& + & \Delta_{n+1} \langle \exp[i \varphi_n(z)-i\varphi_{n+1}(z)]
\rangle a_{n+1}+V_n a_n, \nonumber
\end{eqnarray}
where $\langle ... \rangle$ denotes the average with respect to $z$
over the spatial oscillation period $\Lambda$. Let us then assume
that:\\
(i) $\beta_n(z)=\rho_n \beta(z)$ and $\gamma_n(z)=\rho_n \gamma(z)$,
where $\rho_n$ can take the values $0$ or $1$. This means that some
waveguides in the array are not modulated (those such that
$\rho_n=0$), whereas the modulated waveguides (those with
$\rho_n=1$)  have the same modulation profiles of loss/gain and
propagation constant, defined by the two real-valued functions
$\gamma(z)$ and $\beta(z)$, respectively.\\
(ii) The modulation functions $\beta(z)$ and $\gamma(z)$ are chosen
such that
\begin{equation}
\langle \exp[ i \varphi(z) ] \rangle=\langle \exp[ -i \varphi(z) ]
\rangle=i \Gamma,
\end{equation}
where $\varphi(z)$ is defined by Eq.(C3) with $\beta_n=\beta(z)$
and $\gamma_n=\gamma(z)$, and $\Gamma$ is a real-valued constant.\\
Under such assumptions, Eqs.(C4) reduce to the following ones
\begin{equation}
i \frac{d a_n}{dz} =  \kappa_{n} a_{n-1}+\kappa_{n+1}a_{n+1}+V_n
a_n,
\end{equation}
where
\begin{equation}
\kappa_n  = \left\{ \begin{array} {cc} \Delta_n & {\rm if} \; \;
\rho_{n-1}=\rho_n
\\
i \Gamma \Delta_n & {\rm if} \; \; \rho_{n-1} \neq \rho_n .
\end{array}
\right.
\end{equation}
and $\Gamma$ is defined by Eq.(C5). In this way, Eqs.(C6) describe
the dynamics in a tight-binding lattice with hopping amplitudes
$\kappa_n$ between adjacent sites $|n\rangle$ and $|n-1 \rangle$
which can assume either real values (when the waveguides $n$ and
$n-1$ are both modulated or both not modulated) or purely imaginary
values (when one of the two waveguides $n$ or $n-1$ is modulated,
but the other it is not). The examples of reflectionless
non-Hermitian lattices discussed in Secs.III.B and IV belong to such
a class of lattices. It should be noted that satisfaction of Eq.(C5)
requires a proper choice of the modulation amplitudes for loss/gain
and propagation constant profiles. For instance, let us assume a
sinusoidal modulation
\begin{equation}
\gamma(z)=A_{\gamma} \cos(2 \pi z/ \Lambda) \; , \; \;
\beta(z)=A_{\beta} \cos(2 \pi z / \Lambda).
\end{equation}
In this case, from Eqs.(C3) and (C5)  one obtains
\begin{equation}
i \Gamma = J_0 \left( \frac{\Lambda (A_{\beta}-iA_{\gamma})}{2 \pi}
\right)
\end{equation}
and hence the amplitudes $A_{\beta}$ and $A_{\gamma}$ must be chosen
in such a way that the zero-order Bessel function $J_0$ at the
complex argument $\Lambda (A_{\beta}+iA_g)/(2 \pi)$ gives a purely
imaginary value. There are several possibilities to satisfy such a
condition; for instance, one could fix the product $A_{\beta}
\Lambda$ and determine, correspondingly, the product $A_{\gamma}
\Lambda$; for example, a choice can be
\begin{equation}
\frac{ \Lambda A_{\beta}}{2 \pi} \simeq 2 \; \; , \; \;
\frac{\Lambda A_{\gamma}}{2 \pi} \simeq 2.096
\end{equation}
which yields $\Gamma \simeq 1.941$.


\begin{thebibliography}{31}
\bibitem{notegroup}
For $\kappa>0$ and $0<q<\pi$, an asymptotic plane wave $\psi_n \sim
\exp(-iqn)$ in the lattice propagates with a group velocity
$v_g=-(dE/dq)=2 \kappa \sin(q)>0$, and it is thus a progressive
wave; conversely, a plane wave $\psi_n \sim \exp(iqn)$ propagates in
the lattice with the group velocity $-2 \kappa \sin(q)$, and it is
thus a regressive wave.
\bibitem{note3}
The reflectionless lattices proposed in Ref.\cite{Sukhorukov} are a
special case of a more general class of reflectionless potentials
for the discrete Schr\"{o}dinger equation introduced in the
mathematic literature (see \cite{dis1,dis2}).
\end{thebibliography}

\end{document}